\documentclass[sigconf]{acmart}

\usepackage{listings}
\usepackage{tabularx}
\usepackage{float}
\usepackage{framed}

\usepackage{booktabs}
\usepackage{longtable}
\usepackage{xspace}

\usepackage{graphicx}
\usepackage{textcomp}
\usepackage{algorithm}
\usepackage{algpseudocode}
\usepackage{xcolor}
\usepackage{url}
\usepackage{enumitem}
\usepackage{comment}
\usepackage{array}

\usepackage{multirow}
\usepackage{makecell}
\usepackage{afterpage}
\usepackage{float}
\usepackage{graphicx}
\usepackage{tabularx}
\usepackage{hyperref}
\usepackage{longtable}
\usepackage{multirow}
\usepackage{array}
\usepackage[justification=centering]{caption}
\usepackage{tikz}
\usetikzlibrary{positioning}
\usepackage{soul}
\usepackage{hyperref}
\usepackage{tablefootnote}

\definecolor{shadecolor}{HTML}{ffffcc}

\definecolor{source}{gray}{0.9}

\usepackage[all]{nowidow}

\copyrightyear{2024}
\acmYear{2024}
\setcopyright{rightsretained}
\acmConference[EASE 2024]{28th International Conference on Evaluation and Assessment in Software Engineering}{June 18--21, 2024}{Salerno, Italy}
\acmBooktitle{28th International Conference on Evaluation and Assessment in Software Engineering (EASE 2024), June 18--21, 2024, Salerno, Italy}\acmDOI{10.1145/3661167.3661204}
\acmISBN{979-8-4007-1701-7/24/06}

\setcopyright{none}
\acmDOI{10.1145/xxx}
\usepackage{draftwatermark}
\usepackage{transparent}

\SetWatermarkText{\transparent{0.5}\fontsize{70}{74}\selectfont \parbox[c][3cm][c]{\paperwidth}{\centering Preprint Version\\EASE 2024}}

\SetWatermarkScale{1}

\begin{document}

\title{Mining REST APIs for Potential Mass Assignment Vulnerabilities}

\author{Arash Mazidi}
\affiliation{
  \institution{Technische Universität Clausthal}
  \country{Germany}
}
\email{arash.mazidi@tu-clausthal.de}

\author{Davide Corradini}
\orcid{0009-0009-2594-4562}
\affiliation{
  \institution{
  University of Verona}
  \country{Italy}
}
\email{davide.corradini@univr.it}

\author{Mohammad Ghafari}
\orcid{0000-0002-1986-9668}
\affiliation{
  \institution{Technische Universität Clausthal}
  \country{Germany}
}
\email{mohammad.ghafari@tu-clausthal.de}

\begin{abstract}

REST APIs have a pivotal role in accessing protected resources.
Despite the availability of security testing tools, mass assignment vulnerabilities are common in REST APIs, leading to unauthorized manipulation of sensitive data.
We propose a lightweight approach to mine the REST API specifications and identify operations and attributes that are prone to mass assignment. 
We conducted a preliminary study on 100 APIs and found 25 prone to this vulnerability.
We confirmed nine real vulnerable operations in six APIs.

\end{abstract}




\maketitle

\section{Introduction}

% \dc{Check citation numbers, they do not start from 1}

REST APIs enable seamless data exchange and functionality integration across different systems. 
The pivotal role of these APIs in today's software industry has made them an attractive target for attackers. For instance, a recent API vulnerability disclosed 1.8 million user accounts from an insurance company ~\cite{attack1milion}. Additionally, a security breach in the AWS S3 bucket of a digital scheduling platform exposed the personally identifiable information (PII) of 3.7 million user accounts~\cite{attack3milion}. Furthermore, a major social media platform reported a breach in its API from late 2021 into 2022, revealing the PII of 5.4 million user accounts. The vulnerability originated from an API designed to help users in finding others~\cite{attack5milion}.

% Speaker recognition, widely used as a biometric authentication or identification mechanism in our daily lives, faces significant security concerns, as demonstrated by recent adversarial attacks~\cite{Guangke2021}. \dc{this last example is not relevant}
%

% API security testing is paramount to ensure secure and resilient data protection. State-of-the-art testing tools designed for REST APIs employ advanced methodologies, capturing real-time API traffic and subsequently analyzing, fuzzing, and replaying this traffic \dc{Other tools (e.g., RestTestGen) will generate traffic from scratch}. The primary objective of these tools is to uncover potential vulnerabilities within APIs, contributing to a comprehensive and proactive approach to API security ~\cite{fuzz1, fuzz2, fuzz3}.

 %

Mass assignment is a critical but overlooked vulnerability in REST APIs. 
It occurs when 
REST APIs allow the unintended modification of attributes, leading to unauthorized manipulation of sensitive data.
This vulnerability arises due to an incorrect configuration of widely used REST API frameworks that typically facilitate automatic binding between input data fields and the internal data representation, such as database columns.

\newpage

The support for identifying mass assignment vulnerabilities in REST APIs is limited. Akto~\cite{Aktotool} and RestTestGen~\cite{RTGtool} are two examples of tools for detecting mass assignment vulnerabilities in REST APIs. 
RestTestGen is an automated black-box testing tool, and Akto is semi-automated.
Nonetheless, existing tools mostly evaluate a \emph{running} API, support to uncover mass assignment vulnerabilities in earlier development stages is limited.

We present LightMass, a tool for mining API endpoints and attributes prone to mass assignment vulnerabilities in REST APIs. 
Unlike existing tools that interact with a running API, LightMass merely relies on the API specification; therefore, it draws developers' attention to potential mass assignment vulnerabilities as early as the API's specification is known.
In particular, 
LightMass inspects operations that handle similar sets of attributes (assuming they handle the same data model), and compares the attributes that a \texttt{GET} operation read and those that a \texttt{POST}, \texttt{PUT}, or \texttt{PATCH} operation writes to.
When a \texttt{GET} operation has more attributes than the other operation (i.e., \texttt{POST}, \texttt{PUT}, or \texttt{PATCH}), the attributes that are only present in the \texttt{GET} operation are considered to be \emph{read-only}, and therefore, these attributes are candidates for mass assignment vulnerabilities.

We conducted a preliminary study on 100 APIs and found 25 candidate APIs (115 endpoints and 133 operations) prone to mass assignment vulnerabilities.
% In particular, we discovered 495 potential attributes for mass assignment spread among 115 endpoints and 133 operations. 
%
We examined potential vulnerabilities in six APIs for which we could access the source code and confirmed the presence of nine vulnerable operations.
In summary, LightMass identifies operations that fulfill the necessary conditions for mass assignment vulnerabilities for later in-depth analysis. 
The fast and simple nature of its approach is helpful in several scenarios, such as 
(i) steering code reviewers' focus on potential issues; (ii) enabling tools such as Akto to perform automated testing of mass assignment vulnerabilities;
and (iii) mining API specifications at large and estimating the potential for mass assignment vulnerabilities in the wild.
LightMass is open-source and publicly available on GitHub.\footnote{\url{https://github.com/arash-mazidi/LightMass}}

The rest of this paper is organized as follows. 
We provide background information about RESTful APIs and the mass assignment vulnerability in Section~\ref{sec:background}. 
We introduce our approach to identify potential mass assignment vulnerabilities in Section~\ref{sec:approach}.
We present our evaluation in Section~\ref{sec:evaluation}.
% We provide a short discussion in Section~\ref{sec:discussion}.
We present related work in Section~\ref{sec:relatedwork}, and conclude the paper in Section~\ref{sec:conclusion}.

\section{Background}
\label{sec:background}

This section introduces REST APIs and the OpenAPI standard for writing API specifications. Subsequently, we introduce the mass assignment vulnerability.

\subsection{REST APIs and OpenAPI specifications}

REST APIs are web APIs that adhere to the REST (REpresentational State Transfer) architectural style. 
They offer a consistent interface for creating, reading, updating, and deleting resources. 
HTTP URIs identify resources, and operations on resources are typically associated with HTTP methods such as \texttt{POST}, \texttt{GET}, \texttt{PUT} (or \texttt{PATCH}), and \texttt{DELETE} to, respectively, \textit{create}, \textit{read}, \textit{update} or \textit{delete} resources.

Developers usually follow the OpenAPI standard to describe the API's structure and behavior.
In particular, the OpenAPI specification file, typically structured in JSON or YAML format, describes the endpoints, operations and their attributes, as well as the request and response schemas.

Listing~\ref{lst:example} presents a snippet of the OpenAPI specification for a Task Management API. 
Following an initial header specifying versions, licenses, and the API's base URL, this specification features an array of paths representing the available URI endpoints in the API. In this example, the HTTP URI leading to a task resource is \texttt{/tasks} (line 12), and the HTTP operations \texttt{GET} and \texttt{POST} (lines 13 and 28) are utilized to retrieve the list of existing tasks and create a new task in the system, respectively. These operations have common attributes such as \texttt{title} (lines 22 and 35) and \texttt{assignee} (lines 24 and 37), which delineate the task's title and the person responsible for it.

\begin{lstlisting}[caption={Excerpt of specification for a REST API},label=lst:example,numbers=left,float=t]
openapi: 3.0.1
info:
  title: Task Management
  description: Task retrieving, creating, and so on.
  version: 1.0.0
  license:
    name: Creative Commons Attribution 3.0
    url: http://creativecommons.org/licenses/by/3.0/
servers:
  url: http://localhost:8080
paths: 
  /tasks: 
    get: 
      summary: Get All Tasks
      responses: 
        "200": 
          description: Successful response
          content: 
            application/json: 
              schema: 
                properties: 
                  title: 
                    type: string
                  assignee: 
                    type: string
                  status: 
                    type: "boolean    
    post: 
      summary: Create a Task
      requestBody: 
        content:
          application/json:
            schema:
              properties:
                title: 
                  type: string
                assignee: 
                  type: string
      responses: 
        "201": 
          description: Created successfully
\end{lstlisting}

\subsection{Mass Assignment Vulnerability}

Developers usually rely on frameworks to build REST APIs. These frameworks, such as Spring for Java, Flask for Python, Express.js for JavaScript, and Laravel for PHP, offer a suite of reusable components and features to facilitate REST API development.
One of the features typically provided by these frameworks is called \textit{auto-binding}, a mechanism that adopts naming conventions to automatically map input data in HTTP requests (i.e., parameters) to the backend data objects (e.g., database columns) when they share the same name. This feature is typically enabled by default for all attributes. 
A \textit{mass assignment vulnerability}, also known as ``object injection'' or ``auto-binding vulnerability'', occurs when developers neglect to disable this feature for attributes that are meant to be read-only. 
Therefore, an attacker can add an extra attribute to an HTTP request (one that was not intended to be changed), and the auto-binding feature would automatically link that attribute to its corresponding internal representation, for example, a database column.
In principle, this attribute is neither part of the API specification nor the API documentation, and it should have not been processed. 
Nevertheless, the attacker who exploits this feature, will be able to manipulate and alter data in the database, posing a significant security risk.

For instance, consider the specification of the Task Management API shown in Listing~\ref{lst:example}.
Suppose the JSON \texttt{task} object within an HTTP request maps to the \texttt{tasks} table in the database, which includes a (supposedly) read-only boolean column named \texttt{status} and two modifiable columns, namely \texttt{title} and \texttt{assignee}.
If the REST framework lacks a proper configuration, it may automatically link an additional \texttt{status} attribute in a ``create task'' request to the corresponding \texttt{status} column in the \texttt{tasks} table. 
That is, an attacker could manipulate a request body of the \texttt{POST\,/tasks} operation by introducing an extra \texttt{status} attribute not specified in the OpenAPI specification. The framework would then automatically associate this additional attribute with the \texttt{status} column in the \texttt{tasks} table, allowing the attacker to overwrite the legitimate value in the database with the manipulated HTTP attribute value.

To prevent mass assignment, developers should blacklist read-only attributes from being auto-bound to the internal data representation of the API.

\section{\NoCaseChange{LightMass}}
\label{sec:approach}

\begin{figure*}
\includegraphics[scale=0.63]
% [width=\textwidth]
{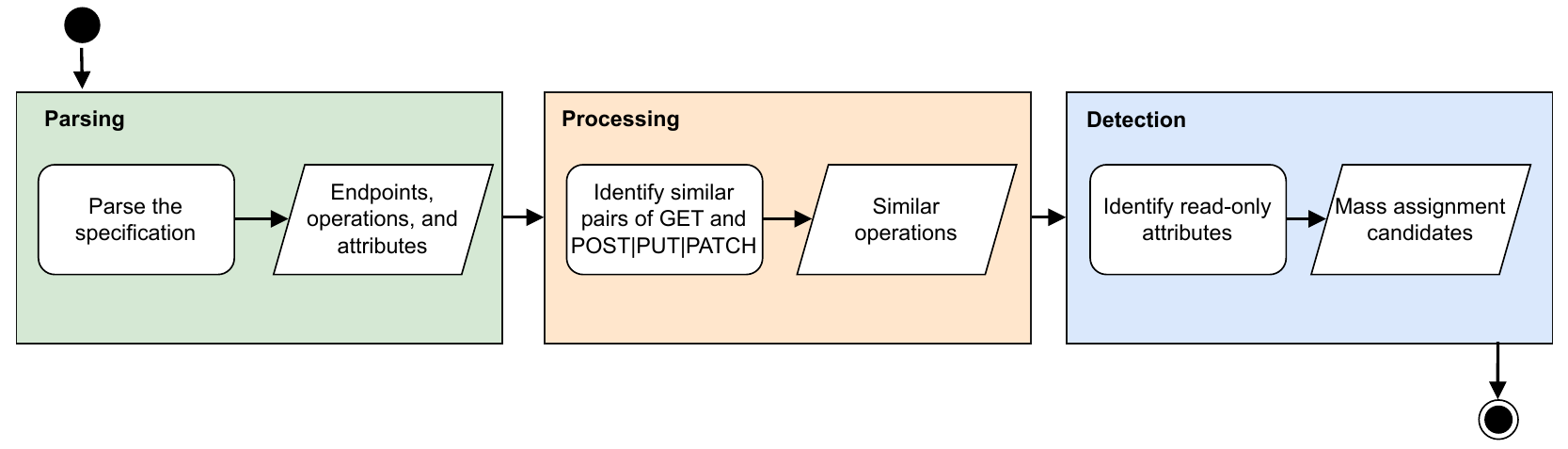}
\caption{LightMass workflow}
\label{fig:workflow}
\end{figure*}

We developed LightMass, a tool that takes an OpenAPI specification file as input and identifies candidate operations and attributes prone to mass assignment vulnerabilities.
Figure~\ref{fig:workflow} illustrates the LightMass workflow, and
Listing~\ref{lst:LightMass} shows the corresponding procedure.

% \begin{figure}
% \includegraphics[width=\columnwidth,trim={0 0.7cm 0 0}]{Figures/LightMass1.pdf}
% \caption{LightMass workflow}
% \label{fig:workflow}
% \end{figure}

\begin{lstlisting}[caption={The procedure to find mass assignment candidates}, label=lst:LightMass,numbers=left,float=t]
Procedure: LightMass
 Input: OpenAPI specification
 Output: Candidate operations and attributes for mass assignment
 
 POST-PUT-PATCH (*$\leftarrow$*) {}
 GET (*$\leftarrow$*) {}
 MassList (*$\leftarrow$*) {}
 
 For Each EndPoint in Specification.Endpoints
    POST-PUT-PATCH (*$\leftarrow$*) EndPoint.POST|PUT|PATCH
    GET (*$\leftarrow$*) EndPoint.GET
 
 FOR Each X in GET
    RES (*$\leftarrow$*) X.RESPOSE.Attributes
    FOR Y in POST-PUT-PATCH
        REQ (*$\leftarrow$*) Y.REQUEST.Attributes

        IF |RES|>|REQ|
            IF RES and REQ are Similar
                MassList (*$\leftarrow$*) (Y, RES - (RES (*$\cap$*) REQ) )
        
 Return MassList
 
EndProcedure
\end{lstlisting}

LightMass parses the API specification to identify existing endpoints, operations, and attributes. It relies on the Jackson library to parse the content. Subsequently, it resolves all the cross references~(i.e., \texttt{\$ref}),\footnote{In OpenAPI, one can define a component at one location in the specification document and reference (reuse) it in other places, reducing redundancy and making the document more maintainable.} ensuring that they are replaced with their actual definitions.
Then, it navigates through the specification to access information about paths and operations (Listing~\ref{lst:LightMass}, lines 9-11). 
The paths object contains details about each endpoint, and under each path, the supported HTTP operations, e.g., \texttt{GET}, \texttt{POST}, \texttt{PUT}, and \texttt{PATCH} are listed.
LightMass extracts all attributes for each operation, which are found within both the request and response bodies, as well as other locations such as path, query, and header (lines 14 and 16). This phase is pivotal since mass assignment vulnerabilities often revolve around the manipulation of input attributes.

LightMass identifies similar operations based on similar attributes.
Firstly, to facilitate a uniform comparison, Porter's stemming algorithm~\cite{porter1980algorithm} is employed to standardize attribute names, i.e., reducing attribute names to their core or root forms.
Secondly, it utilizes the Jaccard coefficient to identify similar operations:
\[
JaccSim(OP, GET) = \frac{|OP.REQ \cap GET.RES|}{|OP.REQ \cup GET.RES|}
\]

Therefore, the similarity measure is the ratio of the number of shared (similar) attributes between two operations to the total number of their distinct attributes (line 20).
Specifically, $OP.REQ$ comprises the attributes in the request body of a \texttt{POST}, \texttt{PUT}, or \texttt{PATCH} operation, and $GET.RES$ is the set of attributes in the response body of a \texttt{GET} operation.

LightMass reports a potential vulnerability when (i) the similarity between two operations is at least 50\%,\footnote{In practice, a vulnerability can exist even with just one extra attribute, but in our experience, a 50\% threshold was practical to uncover actual vulnerabilities and avoid false positives. Nonetheless, it is possible to adjust the similarity threshold in each run if needed.} and (ii) the number of attributes in the response of a \texttt{GET} operation exceeds the number of attributes in the request of the other operation (\texttt{POST}, \texttt{PUT}, or \texttt{PATCH}).
The additional attributes in the \texttt{GET} operation are supposed to be read-only, making them potential candidates for mass assignment vulnerabilities.
In the end, LightMass provides a structured list of candidate endpoints, operations, and attributes prone to mass assignment vulnerabilities. 

For instance, consider the API specification in 
Listing~\ref{lst:example}.
With two attributes (\texttt{title} and \texttt{assignee} at lines 35 and 37) in the request body of the \texttt{POST} operation (line 28) and three attributes (\texttt{title}, \texttt{assignee}, and \texttt{status} at lines 22, 24, and 26) in the response body of the \texttt{GET} operation (line 13), the Jaccard similarity score is 0.66. 
The number of attributes in the \texttt{GET} operation exceeds that of the \texttt{POST} operation.
Therefore, the \texttt{status} attribute in the response body, which is absent in the request body, is a candidate for mass assignment vulnerability.

% \boxit[yellow!20]{
\begin{shaded*}
It is noteworthy that an actual vulnerability exists only if enough protection measures are not in place. Therefore, LightMass' report serves as a guide for security analysts, developers, and testers in conducting further investigations. For example, they could verify that they have properly disabled the auto-binding feature for the attributes flagged as potentially vulnerable.
\end{shaded*}

\section{Evaluation}
\label{sec:evaluation}

\begin{table*}[t]
\centering
\begin{tabular}{l|c|c|c|c|c}
\toprule
 % &  &
 & \multicolumn{2}{c|}{\textbf{Total}}
 &\multicolumn{3}{c}{\textbf{Flagged vulnerable}}\\
\textbf{API Name} & \textbf{\# Endpoints} & \textbf{\# Operations} & \textbf{\# Endpoints} & \textbf{\# Operations} & \textbf{\# Attributes} \\
\midrule
VAmPI & 10 &  12 & 2 & 2 & 2 \\
OWASP & 4 &  10 & 2 & 2 &  4 \\
Toggle & 8 &   16 & 2 & 2 & 2 \\
CRUD & 1 &   4 & 1 & 2 & 2 \\
Bookstore & 3 &  5 & 1 & 1 & 1 \\
StudentAPI & 5 &  5 & 2 & 2 & 2 \\
\midrule
Search Console & 7 &   11 & 1 & 1 &  2 \\
Fitness & 7 &  13 & 4 & 4 &  7\\
Calendar & 22 &   37 & 7 & 10 & 18 \\
My Business & 40 &   50 & 9 & 9 & 22 \\
Analytics & 43 &   88 & 22 & 28 & 175 \\
Classroom & 34 &   61 & 17 & 17 & 27 \\
YouTube & 39 &   76 & 12 & 20 & 106 \\
SpaceX & 52 &   94 & 1 & 1 & 3 \\
Reservations & 5 &   7 & 1 & 1 & 1 \\
ProjectManagement & 58 &   78 & 1 & 1 & 1 \\
AlerterSystem & 186 &   422 & 4 & 4 &  30 \\
TransferService & 3 &   3  & 1 & 1 &  16 \\
CheckoutService & 23 &   24 & 1 & 1 &  3 \\
Registry & 20 &   35  & 10 & 10 &  10 \\
SMS & 2 &  5  & 2 & 2 &  24 \\
ATS & 4 &   5 & 1 & 1 &  13 \\
Auto Scaling & 65 &  130 & 1 & 1 &  1 \\
Docker HUB & 20 &   26 & 1 & 1 &  3 \\
Files & 134 &   222 & 9 & 9 &  20 \\
\midrule
\textbf{Total} & \textbf{795} &   \textbf{1439} & \textbf{115} & \textbf{133} &  \textbf{495} \\
\bottomrule
\end{tabular}
\caption{LightMass report for 25 REST APIs}
% \dc{add columns of total attributes, if there is time}
\label{tab:output}
\end{table*}

% \subsection{Result}

We applied LightMass to 100 APIs that we randomly collected from previous work~\cite{RTG2023}, GitHub, the Google APIs, APIs Guru,\footnote{\url{https://apis.guru/}} and EMB.\footnote{\url{https://github.com/EMResearch/EMB}}

% \newpage
Mining the OpenAPI specifications of these APIs uncovered 25 potentially vulnerable APIs listed in Table~\ref{tab:output}.
% , comprising 795 endpoints and 1439 operations.
Specifically, LightMass reported 495 candidate attributes for mass assignment distributed across 115 endpoints and 133 operations in 25 APIs.
%

% \newpage

% This result indicates that 14\% of the total endpoints (115 out of 795) and 9\% of the total operations (133 out of 1439) exhibit characteristics that make them susceptible to mass assignment vulnerabilities.

% These results underscore the prevalence of potential vulnerabilities in API implementations and highlight the importance of early detection and mitigation strategies in the software development process. 
% We have summarized these findings in Table~\ref{tab:results2}, which provides a clear and concise overview of LightMass's performance across different APIs.

\begin{comment}
\begin{table}
\centering
\caption{Potential vulnerable REST APIs}
\begin{tabular}{|l|c|c|c|}
\hline
API Name & \# Endpoints & \# Operations & \# Attributes \\
\hline
VAmPI & 2 & 2 & 2 \\
OWASP & 2 & 2 &  4 \\
Toggle & 2 & 2 & 2 \\
CRUD & 1 & 2 & 2 \\
Bookstore & 1 & 1 & 1 \\
StudentAPI & 2 & 2 & 2 \\
\hline
Search Console & 1 & 1 &  2 \\
Fitness & 4 & 4 &  7\\
Calendar & 7 & 10 & 18 \\
My Business & 9 & 9 & 22 \\
Analytics & 22 & 28 & 175 \\
Classroom & 17 & 17 & 27 \\
Youtube & 12 & 20 & 106 \\
spacex-api & 1 & 1 & 3 \\
reservations-api & 1 & 1 & 1 \\
projectManagement & 1 & 1 & 1 \\
alertersystem & 4 & 4 &  30 \\
TransferService & 1 & 1 &  16 \\
CheckoutService & 1 & 1 &  3 \\
registry & 10 & 10 &  10 \\
sms & 2 & 2 &  24 \\
ats & 1 & 1 &  13 \\
autoscaling & 1 & 1 &  1 \\
hub & 1 & 1 &  3 \\
files & 9 & 9 &  20 \\
\hline
Total & 115 & 133 &  495 \\
\hline
\end{tabular}
\label{tab:results2}
\end{table}

\end{comment}

Unfortunately,
there is no golden dataset for mass assignment vulnerabilities in REST APIs.
To evaluate whether these APIs are actually vulnerable, we should either test the APIs or examine their code. It is unethical to test APIs in production due to the potential risk of launching a successful attack. Hence, we compared LightMass and existing tools against six open-source APIs that we could set up and run locally.
These APIs are listed in Table~\ref{tab:open-sourceAPIs}.

% To evaluate whether these endpoints are actually vulnerable, we checked six APIs, listed in Table~\ref{tab:open-sourceAPIs}, for which we had access to the source code and could run them locally.
%
% We could not test active APIs due to ethical considerations. 
% This precaution aimed to prevent the potential risk of launching a successful attack on APIs in production.

\begin{table*}[t]
% \centering
\begin{tabular}{p{2cm}|p{1.3cm}|p{12cm}}
\toprule
\textbf{API Name} & \textbf{Language} & \textbf{Description} \\
\midrule
VAmPI ~\cite{vampi} &  Python &  A vulnerable API which includes all the OWASP top 10 vulnerabilities of APIs.\\
\midrule
OWASP~\cite{OWASP} &  Java & An API vulnerable to broken object-level authorization, excessive data exposure, and mass assignment.\\
\midrule
Toggle~\cite{Toggle} & ASP.Net & It defines toggles for a list of services.\\
\midrule
CRUD~\cite{CRUD} & Node.js & A CRUD example with NodeJS, Sequelize, Swagger, and MySQL.\\
\midrule
Bookstore~\cite{Bookstore} & Java & An API designed to expose the features to manage a book store.\\
\midrule
StudentAPI~\cite{StudentAPI} & Java & It is a vulnerable API intended for educational purposes, with a focus on addressing mass assignment vulnerabilities.\\
\bottomrule
\end{tabular}
\caption{The open-source APIs used as case studies}
\label{tab:open-sourceAPIs}
\end{table*}
%

% \newpage
To identify existing tools for mass assignment detection and compare them with LightMass, 
we searched the literature and Google with a combination of keywords such as \textit{mass assignment} and \textit{detection, scanner, or analyzer}. 
We also extended our search to GitHub with keywords such as \textit{mass assignment}, \textit{object injection}, and \textit{autobinding}.
Upon obtaining a list of potential tools, we paid close attention to the repository descriptions, README files, and any available documentation to determine the relevance of every search result.

We identified a total of nine (semi-)automated tools. 
We eliminated two since they had not been updated since 2010, suggesting potential obsolescence. 
We scrutinized the remaining tools and discovered that five are designed for mass assignment detection in web applications. 
The two remaining tools, namely RestTestGen~\cite{RTGtool} and Akto~\cite{Aktotool}, supported mass assignment detection in REST APIs.

Akto cannot automatically identify mass assignment vulnerabilities, so we had to manually input the potential vulnerable endpoints and attributes.\footnote{We provided Akto with the output from LightMass and checked whether Akto flags them for mass assignment vulnerability or not.
}
Therefore, we applied RestTestGen to the six APIs in Table~\ref{tab:open-sourceAPIs} to build our ground truth for mass assignment vulnerabilities.

\begin{table}[t]
\begin{tabular}{l|c|c|c}
\toprule
\textbf{API Name} & \textbf{Akto} & \textbf{RestTestGen} & \textbf{LightMass} \\
\midrule
VAmPI & 1 & 1 &  2 \\
OWASP & 4 & 4 & 4 \\
Toggle & 2 & 2 &  2 \\
CRUD & 2 & 2 &  2  \\
Bookstore & 1 & 1 &  1 \\
StudentAPI & 2 & 2 &  2 \\
\bottomrule
\end{tabular}
\caption{The number of attributes flagged by each tool}
\label{tab:manualresults}
\end{table}

Table~\ref{tab:manualresults} lists the vulnerability reports by each tool.
Akto and RestTestGen provided the same results, whereas LightMass flagged one more attribute prone to mass assignment vulnerability in the VAmPI API.
We looked at the source code of VAmPI to learn about the extra attribute (named \texttt{owner}) that LightMass had flagged.  
Upon inspection, we found that the \textit{book} model in the VAmPI API had predefined fields allowed to be set while creating a new book instance, namely \texttt{book\_title}, \texttt{secret\_content}, and \texttt{user\_id}. 
These were the only permissible fields for setting while creating a new book. 
Any attempt to include an additional field in the request, such as \texttt{owner}, would be unsuccessful due to a server-side restriction and input validation. 
Therefore, 
the extra field that LightMass flagged for VAmPI API was a false positive.

It is important to note that the obtained results for these six
case studies cannot be generalized to the remaining 19 (unverified) APIs.

In summary, the preliminary evaluation results are promising. 
Nonetheless, relying on LightMass as a standalone tool requires future studies. 
Particularly, how it performs in terms of false positives against APIs that are not vulnerable remains for a future work investigation.
It is noteworthy that an actual
vulnerability exists only if enough protection measures are not in place.
Hence, relying merely on the specification is not enough, and false positives are expected.
Nonetheless, there is no tool support for early development stages in this domain, and we believe that LightMass draws developers' attention to this overlooked problem.
In addition, as we experimented, LightMass enables Akto to act as a fully automated testing tool for mass assignment vulnerabilities.
Akto is a popular and comprehensive API testing tool that does not support automated testing for mass assignment. It requires human interventions and input for suspected operations and attributes.
Hence, LightMass in its current state enables the community to apply Akto as a fully automated tool to uncover true mass assignment vulnerabilities. 

\section{Related Work}\label{sec:relatedwork}

% REST APIs play a pivotal role in today's software industry, and researchers have put great effort into exploring the potential vulnerabilities in these APIs and their ecosystems.
%

Gadient et al.~\cite{Gadient2020} mined 9,714 Web APIs from 3,376 mobile apps, and found that in 500 apps, these APIs transmit embedded code (e.g., SQL and JavaScrip commands), exposing the app users and web servers to code injection attacks.
% They also found hardcoded API keys
%
In a follow-up study~\cite{Gadient2021}, they also discovered that API servers are usually misconfigured.
They observed that on average every second server suffers from version information leaks, and worryingly, servers are typically set up once and then left untouched for up to fourteen months, yielding severe security risks.

Atlidakis et al.~\cite{Atlidakis2019} presented RESTler, a stateful REST API fuzzer that examines the OpenAPI specification. RESTler statically analyzes OpenAPI specification and creates and executes tests by deducing dependencies and examining the responses from previous test runs.
They also demonstrated an extension of RESTler with active property checkers, which enables automatic testing and identification of breaches in adherence to these rules~\cite{Atlidakis20200}.
Godefroid et al.~\cite{Godefroid2020} conducted a study on the intelligent generation of data payloads within REST API requests, leveraging the OpenAPI specification. 
They showed that they can detect data-processing vulnerabilities in cloud services.
They~\cite{Godefroid2018} also presented a method for automated differential regression testing of REST APIs aimed at identifying breaking changes between API versions by comparing the responses of various versions when given the same inputs to identify discrepancies and identifying regressions in these observed differences. 
Mirabella et al.~\cite{Mirabella2021} presented a deep learning model to predict the validity of test inputs in an API request before making the API call. 

Mai et al.~\cite{Mai2019} introduced a tool designed to automatically create executable security test cases from misuse case specifications written in natural language. 
Reddy et al.~\cite{Reddy2022} introduced an approach centered around sequence models and transformers to discern whether an API request is prone to injection attacks.
% contains SQL injections, code injections, XSS attacks, operating system command injections, or other forms of malevolent injections.
%
Barabanov et al.~\cite{Barabanov2022} introduced an automated technique for identifying vulnerable endpoints to  Insecure Direct Object Reference (IDOR) and Broken Object Level Authorization (BOLA) vulnerabilities which are related to the improper handling of object references, particularly in the context of authorization.
This method involves establishing a mapping between attack methodologies and the properties in endpoints found within OpenAPI specifications.

The number of studies on mass assignment vulnerability is limited.
Corradini et al.~\cite{RTG2023} developed an automated black-box testing approach to find mass assignment vulnerabilities in RESTful APIs. 
The approach, built on top of the RestTestGen framework~\cite{corradini2022resttestgen}, uses EM clustering to group operations within the endpoints. Subsequently, abstract testing templates are instantiated to automatically generate interaction sequences, to exploit potential vulnerabilities. This tool requires the API to be in a running status for interaction with HTTP operations.

\newpage
Park et al.\cite{FUGIO2022} introduced an automated tool for generating exploits targeting PHP object injection vulnerabilities named FUGIO. 
Koutroumpouchos et al.~\cite{objectmap2019} introduced ObjectMap, a customizable solution that identifies deserialization and object injection vulnerabilities in web applications using Java and PHP. 
Shcherbakov et al.~\cite{serialdetector2021} introduced SerialDetector, a taint-driven dataflow analysis technique to identify Object Injection Vulnerabilities (OIVs) patterns within .NET assemblies.

% \ins{In summary, LightMass serves as a light-weight tool that guides  security analysts and testers in conducting further investigations and uncovering actual mass assignment  vulnerabilities.}

\section{Conclusion}
\label{sec:conclusion}

Mass assignment is a critical vulnerability in REST APIs.
However, there is a lack of support for developers to identify this security risk in the early stages of API development.
We introduced LightMass, a tool that mines REST API specifications for potential mass assignment vulnerabilities.
It identifies operations and attributes that fulfill the necessary conditions for mass assignment vulnerabilities. 
LightMass is not dependent on an active API, and it can alert developers as soon as the API's specification is known. 
It also enables Akto, the popular open-source API testing tool, to execute fully automated API testing for mass assignment vulnerabilities.
%

% Developing a white-box testing approach to uncover mass assignment vulnerabilities would be a potential future direction. 

% We mined 100 OpenAPI specifications and found 25 potentially vulnerable APIs. 
% We confirmed nine vulnerable endpoints in six open-source APIs.

\bibliographystyle{ACM-Reference-Format}
\bibliography{ease2024-37}

\end{document}